\documentstyle[12pt,psfig]{article}
\begin{document}
\newcommand{\lsim}
{\ \raisebox{2.75pt}{$<$}\hspace{-9.0pt}\raisebox{-2.75pt}{$\sim$}\ }
\newcommand{\gsim}
{\ \raisebox{2.75pt}{$>$}\hspace{-9.0pt}\raisebox{-2.75pt}{$\sim$}\ }
\newcommand{\ber}{\begin{eqnarray}}
\newcommand{\eer}{\end{eqnarray}}
\newcommand{\ind}{{\white em}}
\newcommand{\mind}{{\white empty}}
\newcommand{\vva}{{\bf v}_{\perp\alpha}}
\newcommand{\vxa}{{\bf x}_{\perp\alpha}}
\newcommand{\vxp}{{\bf x}_\perp}
\newcommand{\pa}{p_{\perp\alpha}}
\newcommand{\gfm}{ {\rm GeV/fm}^3}
\newcommand{\beq}{\begin{equation}}
\newcommand{\eeq}[1]{\label{#1} \end{equation}}
\newcommand{\half}{{\textstyle \frac{1}{2}}}
\newcommand{\gton}{\stackrel{>}{\sim}}
\newcommand{\lton}{\mathrel{\lower.9ex\hbox{$\stackrel{\displaystyle 
<}{\sim}$}}}
\newcommand{\ee}{\end{equation}} \newcommand{\ben}{\begin{enumerate}}
\newcommand{\een}{\end{enumerate}} \newcommand{\bit}{\begin{itemize}}
\newcommand{\eit}{\end{itemize}} \newcommand{\bc}{\begin{center}}
\newcommand{\ec}{\end{center}} \newcommand{\bea}{\begin{eqnarray}}
\newcommand{\eea}{\end{eqnarray}}
\newcommand{\beqar}{\begin{eqnarray}}
\newcommand{\eeqar}[1]{\label{#1} \end{eqnarray}}
\newcommand{\vx}{{\bf x}}
\newcommand{\vp}{{\bf p}}
\newcommand{\vpa}{{\bf p}_{\perp\alpha}}
\newcommand{\vpp}{{\bf p}_\perp}
\newcommand{\vkp}{{\bf k}_\perp}
\newcommand{\vqp}{{\bf q}_\perp}
\newcommand{\ya}{y_\alpha}
\newcommand{\sh}{{\rm sh}}
\newcommand{\ch}{{\rm ch}}
\newcommand{\de}{\sigma}
\newcommand{\ep}{{\rm e}_\perp}
\newcommand{\call}{{ L}}

\begin{flushright}
DOE/ER/40561-321-INT97-21-05\\
CU-TH-826\\
TPI-MINN-97-07\\[4ex]
\end{flushright}

\begin{center}
  {\Large Yang-Mills Radiation in Ultra-relativistic Nuclear Collisions}
  \footnote[0]{PREPARED FOR THE U.S. DEPARTMENT OF ENERGY UNDER GRANT
    DE-FG06-90ER40561\\ This report was prepared as an account of work
    sponsored by the United States Government. Neither the United States nor
    any agency thereof, nor any of their employees, makes any warranty, express
    or implied, or assumes any legal liability or responsibility for the
    accuracy, completeness, or usefulness of any information, apparatus,
    product, or process disclosed, or represents that its use would not
    infringe privately owned rights.  Reference herein to any specific
    commercial product, process, or service by trade name, mark, manufacturer,
    or otherwise, does not necessarily constitute or imply its endorsement,
    recommendation, or favoring by the United States Government or any agency
    thereof. The views and opinions of authors expressed herein do not
    necessarily state or reflect those of the United States Government or any
    agency thereof.} \\[3ex]

{M. Gyulassy$^{1,3}$
 and L. McLerran$^{2,3}$
}\\[2ex]

{\small 
1.Department of Physics, Columbia University, New York, NY  10027  USA\\
2.Physics Department, University of Minnesota, Minneapolis, MN  55455  USA\\
3.Institute for Nuclear Theory, University of Washington, Box 351550
Seattle, WA 98195, USA
}\\[2ex]
{\today}\\[3ex]
\end{center}

Abstract: The classical Yang-Mills radiation computed in the
McLerran-Venugopalan model is shown to be equivalent to the gluon
bremsstrahlung distribution to lowest ($g^6$) order in pQCD.  The classical
distribution is also shown to match smoothly onto the conventional pQCD
mini-jet distribution at a scale $k_\perp^2\sim \chi$, characteristic of the
initial parton transverse density of the system.  The atomic number and energy
dependence of $\chi$ is computed from available structure function information.
The limits of applicability of the classical Yang-Mills description of
nuclear collisions at RHIC and LHC energies are discussed.\\[2ex]


\newpage

\section{Introduction}

In this paper, we compare recent classical and quantal derivations of induced
gluon radiation for applications to ultra-relativistic nuclear collisions.  The
classical distribution, based on the McLerran-Venugopalan model\cite{mcven},
was recently computed to order $g^6$ in Ref.\cite{kmw}.  The soft gluon
bremsstrahlung distribution was computed via pQCD by Bertsch and Gunion in
Ref.\cite{gunion} within the Low-Nussinov approximation.  Another quantal
distribution, based on the Gribov-Levin-Ryzkin ladder
approximation\cite{gribov}, was recently applied by Eskola et al\cite{esk}.
Finally, there has been considerable recent effort to compute moderate
$p_\perp$ (mini-jet) distributions based on the conventional collinear
factorized pQCD approach\cite{mini,wang,geiger}.

Interest in the moderate $p_\perp$ gluon distributions arises in connection
with estimates of the initial conditions and early evolution of the quark-gluon
plasma formed in ultra-relativistic nuclear collisions at RHIC ($\surd s=200$
AGeV) and LHC ($\surd s=6500$ AGeV) energies.  Until recently, the main source
of mid-rapidity gluons was assumed to be copious mini-jet production as
predicted via the conventional pQCD $gg\rightarrow gg$
processes\cite{mini,wang,geiger}.  However, in Refs.\cite{mcven,kmw} it was
suggested that another important source of mid-rapidity gluons could be the
classical Yang-Mills bremsstrahlung associated with the passage of two heavy
nuclei through each other.  In the conventional approach, beam jet
bremsstrahlung is assumed to influence only the non-perturbative low transverse
momentum beam jet regions.  Beam jets are then typically modeled by pair
production in Lund or Dual Parton Model strings. See for example Refs.
\cite{wang,geiger} and references therein.

The novel suggestion in \cite{mcven} was that for sufficiently large $A$ nuclei
and high energy, the initial nuclear parton density per unit area could become
so high that the intrinsic transverse momentum of the partons, $\surd \chi
\propto A^{1/6} \Lambda_{QCD}$, could extend into the mini-jet perturbative
regime, $k_\perp\sim 2-4$ GeV.  It was suggested that beam jet bremsstrahlung
could even dominate that few GeV transverse momentum region because it is
formally of lower order in $\alpha_s$ than mini-jet production.  Such a new
source of moderate $p_\perp$ partons would then significantly modify the early
$\tau\sim 1/\surd \chi$ evolution and hence possibly modify many of the
proposed signatures of the quark-gluon plasma in such reactions\cite{qm96}.
 
One of the aims of the present paper is to show in fact that the classical and
quantal bremsstrahlung and mini-jet sources of mid-rapidity gluons are actually
equivalent up to form factor effects over a continuous range of $p_\perp \sim
\surd \chi$ regime. In addition we explore the limits of the validity of each
approximation and compute numerically the energy and atomic number dependence
of the McLerran-Venugopalan density parameter, $\chi$.  This parameter is the
total color charge squared per unit area of partons with rapidities exceeding
some reference value.

The calculation of this paper checks that there is a region of overlap between
the classical and quantum computation.  The quantum calculation should be valid
at large transverse momenta.  The classical calculation is valid at momenta
$\Lambda_{QCD} \ll k_\perp \ll
 \surd s $.  Most of the gluons are produced in
the region appropriate for the classical calculation.  It is well known that
perturbative calculations of gluon production are power law sensitive to an
infrared cutoff.  The classical computation has this infrared cutoff built into
the calculation and may ultimately lead to a proper computation of gluon
production.  The region where we can compare the calculations is at $k_\perp$
much greater than this cutoff.

The plan of this paper is as follows: In section 2, we review the classical
derivation of induced gluon radiation in the McLerran-Venugopalan model. We
correct the treatment in \cite{kmw} of the contact term in the classical
equations of motion for a single nucleus.  We extend further that derivation to
treat properly the renormalization group corrections to the density parameter
$\chi$.  Those corrections\cite{jamil} increase significantly  the color
charge squared per unit area relative to the contribution of the valence quarks
thusfar considered
\cite{kmw,yuri,rischyuri}. 
 We also correct omitted factors of 2 and $2\pi$ in the
original computation\cite{kmw}.

In the third section, we review pQCD based derivations of induced gluon
radiation\cite{gunion}-\cite{gribov}.  We show that the classical result agrees
with the quantal results of Bertsch and Gunion.\cite{gunion} and also with the
GLR formulation\cite{gribov} if DGLAP evolution of the structure functions is
assumed.  We then compare the bremsstrahlung distribution to the mini jet
distribution and show that while the latter dominates at high transverse
momentum $p_\perp \gg \surd \chi$, the former dominates at $p_\perp\ll \surd
\chi$. However, there is a continuous range of momenta $\sim \surd\chi$ where
both results agree at the level of $20-40\%$.

In the fourth section, we compute the parameter $\chi$ of the
McLerran-Venugopalan model.  We find that due to the rapid rise of the small
$x$ gluon structure functions, $\surd \chi $ approaches on the order of 1 GeV
by LHC energies for $A\sim 200$.  Possible implications and further
extensions of this model  conclude the paper.

\section{Classical Yang-Mills Radiation}

The basic assumption of the classical approach that follows is that
the coupling strength is small at the scale
\beq
        \Lambda^2 = {1 \over {\pi R^2}} {{dN} \over {dy}} \gg \Lambda_{QCD}^2
\; \; . \eeq{lambda}
 The parameter $\Lambda^2$ is 
the number density of gluons per unit rapidity 
per unit area. .  

The gluon distribution was shown in Ref. \cite{jamil} to solve an evolution
equation which in various limits is the BFKL equation\cite{bfkl}, the DGLAP
equation\cite{dglap}, or its non-linear generalization\cite{gribov}.  In the
ultra-relativistic domain, also the rate of change of the multiplicity per unit
rapidity \beq {{d^2N/dy^2} \over {dN/dy}} \sim \alpha_s \eeq{vardndy} is small.
The smallness of this parameter means that if we compute the gluon distribution
in a small region $\Delta y \sim 1$ around $y = 0$, then the source of those
gluons is dominated by hard partons with rapidities much larger than 1. These
hard partons can be integrated out of the effective action which describes the
color field source at $y \sim 1$, and they lead to an effective external
classical static source for the gluon field.

Since this can be done at any reference rapidity, the classical gluon field may
be thought of as arising from a rapidity dependent classical source.  For a
single nucleus moving near the positive light cone, we have \beq D_\mu F^{\mu
  \nu} = g^2 \delta^{\nu +} \rho(x^-, x_\perp) \eeq{eqnmo1} with the source
approximately independent of $x^+=(t+z)/\surd 2$.  Two types of rapidity
variables must be differentiated.  In the classical equation of motion, the
coordinate space rapidity is relevant as defined by \beq y = \ln 1/x^-=
y_{proj} - \ln(x^-/x^-_{proj}) \; \; , \eeq{rap4} and $x^-=(t-z)/\surd 2$.  The
momentum space rapidity is, on the other hand, \beq y = {1 \over 2} \ln
(p^+/p^-) \eeq{rap1} where \beq p^\pm = {1 \over \sqrt{2}}(p^0\pm p^3)
\eeq{rap2} are the conjugate momenta to $x^\pm$.  For a hadron with
$p^+=p^+_{proj}$, we define $x_{proj}^- = exp(-y_{proj}) \sim 1/p^+_{proj}$.

The coordinate space rapidity is of the same order as the momentum space
rapidity, since by the uncertainty principle $ \Delta x^- \sim 1/p^+$.
Qualitatively, these rapidities may thus be thought of as interchangeable.  On
the other hand, the classical equations of motion are described by coordinate
space variables, and we must use the coordinate space rapidity.

In the McLerran-Venugopalan model, the source rapidity density \beq
\rho(y,x_\perp)\equiv x^-\rho(x^-, x_\perp) \; \; ,\eeq{rapsou} is assumed to
be a stochastic variable which is integrated over with a Gaussian weight, \beq
\int [d\rho ] exp\left( - \int dy d^2x_\perp {1 \over \mu^2 (y)} {\rm Tr}\rho^2
(y, x_\perp) \right) \; \; .\eeq{measure} This Gaussian assumption ignores
correlations which we will see later are needed to regulate the infrared
singularities.  Here $\mu^2(y)$ is the average charge squared per unit rapidity
per unit area scaled by $1/(N_c^2-1)$ \beq \mu^2(y) = {1 \over {N_c^2-1}} {1
  \over {\pi R^2}} {{dQ^2} \over {dy}} \eeq{mu2} Note that this $\mu^2(y)$
specifies the rms fluctuations of the charge transverse density at a fixed
rapidity.  The quantity analogous to the rapidity independent $\mu$ used in
\cite{kmw} is the integrated transverse density of color charge arising from
hard partons exceeding a reference rapidity.  To emphasize this distinction we
denote this quantity by \beq \chi (y) = \int_y^{y_{proj}} dy^\prime \mu^2
(y^\prime ) \; \; . \eeq{chi} 
This quantity will related below to the integrated gluon
structure function.

The solution to the above equations may be found in the light cone gauge
by assuming that
\begin{eqnarray}
        A^\pm  =  0 \nonumber \; \; , \; \;
        A^i = A^i (y, x_\perp) 
\end{eqnarray}
The index $i=1,2$ ranges  over only the two-dimensional transverse
coordinates.  The field $A^i$ solves
\beq
        -D_i {d \over {dy}} A^i = g^2  \rho (y,x_\perp)
\; \; . \eeq{aperp}
Equation (\ref{aperp}) is solved by letting\cite{jamil}-\cite{yuri}
\beq
        A^i (y,x_\perp) = {1 \over i} \left( P e^{i \int _{y_{proj}}^y 
dy^\prime
\Lambda (y^\prime, x_\perp )} \right)^\dagger 
\nabla^i \left( P e^{i \int _{y_{proj}}^y dy^\prime
\Lambda (y^\prime, x_\perp )} \right)
\eeq{aperp1}
In this equation, $P$ denotes path ordering along the integration in rapidity.

If we now change variables (with unit  determinant in the integration
over sources)
\beq
        \left( P e^{i \int _{y_{proj}}^y dy^\prime
\Lambda (y^\prime, x_\perp ) }\right) \rho 
\left( P e^{i \int _{y_{proj}}^y dy^\prime
\Lambda (y^\prime, x_\perp )} \right)^\dagger \rightarrow \rho
\;\; ,\eeq{newrho}
then $\Lambda $ is seen to obey the two dimensional Poisson's equation
\beq
        - \nabla^2_\perp \Lambda (y, x_\perp) = g^2 \rho (y,x_\perp)
\eeq{lameqn}

Note that due to the expected slow variation of the source
 density as a function of rapidity, the field
is almost constant in $y$.  At zero rapidity, therefore,
the field may be taken  approximately as
\beq
        A_i(x^-,x_\perp) = \theta (x^- ) \alpha _i^+ (x_\perp)
\eeq{aperpapp}
where
\beq
        \alpha_i^+ (x_\perp) = {1 \over i}
\left( P e^{i \int _{y_{proj}}^0 dy^\prime
\Lambda (y^\prime, x_\perp ) }\right)^\dagger \nabla_i
\left( P e^{i \int _{y_{proj}}^0 dy^\prime
\Lambda (y^\prime, x_\perp ) }\right)
\eeq{alpha}
This is the  non-abelian Weizs\"acker-Williams field of the projectile
nucleus which must still be averaged over the ensemble (\ref{measure}).

In order to generalize the above solution to the case of two colliding nuclei,
we use the same variables as above for the projectile nucleus propagating in
the $+z$ direction.  For the target nucleus propagating in the $-z$ direction,
we use the rapidity variable \beq y = -y_{cm} + \ln(x_0^+/x^+) \eeq{rapnuc2}
Here we denote the projectile rapidity with the center of mass rapidity as,
$y_{cm} = y_{proj}$.  We will also henceforth use the index $+$ to refer to $y
> 0$ and $-$ to $y < 0$, when no confusion will arise with respect to light
cone variable indices.

In the neighborhood of $y = 0$, we can ignore the small
rapidity dependence of the 
fields.  The solution to the equations of motion
in the 
\beq
        x^+ A^- + x^- A^+ = 0
\eeq{gaugecond}
gauge is  approximately given by
\beq
        A^\pm = \pm x^\pm \theta (x^+) \theta (x^-) \beta (\tau, x_\perp)
\eeq{apm}
and
\beq
        A_i = \theta (x^+) \theta (x^-) \alpha_i^3 (\tau, x_\perp) +
\theta (-x^+) \theta (x^-) \alpha^+_i(x_\perp) + \theta (x^+) \theta (-x^-)
\alpha^-_i (x_\perp)
\eeq{alphai}
Here $\tau = \sqrt{t^2 - z^2}$ is a boost covariant time variable.
(Note that the above notation corresponds to $\beta=\alpha$ 
and $\alpha^3_i=\alpha_{i\perp}$ of \cite{kmw}.)

The fields
\beq
        \alpha^\pm_i  = {1 \over i} \left( P e^{i \int_{\pm y_{cm}}^0
dy^\prime \Lambda (y^\prime, x_\perp)} \right)^\dagger \nabla_i
 \left( P e^{i \int_{\pm y_{cm}}^0
dy^\prime \Lambda (y^\prime, x_\perp)} \right)
\eeq{alphapmi}
where
\beq
        -\nabla^2_\perp \Lambda (y, x_\perp) = g^2 \rho (y, x_\perp)
\eeq{freelam}
and where
\beq
        \rho (y, x_\perp) = \theta (y) \rho^+ (y,x_\perp) + \theta (-y) \rho^-
(y,x_\perp)
\eeq{rhopm}

Notice that in this solution, the fields $\alpha_i^\pm$ are two dimensional
gauge transforms of vacuum fields.  Their sum is of course not a gauge
transform of vacuum fields, and therefore the solution cannot continue into the
region $x^\pm > 0$.  There is in fact a singularity in the solution at $x^+ =
0$ and $x^- = 0$, at $x^+ = 0$ for $x^- > 0$, and at $x^- = 0$ for $x^+ > 0$.
For $x^\pm > 0$, the form of the fields chosen above solves the classical
equations of motion.  In this region, the solution is a boundary values problem
with the boundary values specified on the edge of the forward light cone.

To determine these boundary values,
we solve
\beq
        D_\mu F^{\mu \pm} = g^2 \rho
\eeq{pmeqn}
and
\beq
        D_\mu F^{\mu i} = 0
\eeq{ieqn}

First we find the singularities of  Eqn. \ref{ieqn}.  In this equation,
there is a $\delta(x^+) \delta (x^-)$ singularity, that is a singularity
at the tip of the light cone.  The absence of such a singularity requires that
\beq
        \alpha_i^3 \mid_{\tau = 0} = \alpha^+_i (x_\perp) + \alpha^-_i 
(x_\perp)
\eeq{alph3bn}
There are also singularities of the form $\delta (x^\pm )$ for
$x^\mp > 0$.  The absence of these singularities requires $\alpha^3 $
be analytic as $\tau \rightarrow 0$.

The solution for the Eqn. \ref{pmeqn} is a little trickier since there are some
potentially singular contact terms.  It can be shown that if the
fields $\alpha^\pm_i$ are properly smeared in rapidity so that
they really solve the equations of motion in the backwards light cone, then 
all such contact terms disappear.  We find that $\beta $ must be analytic
at $\tau = 0$ and that
\beq
        \beta \mid_{\tau = 0} = {i \over 2} [\alpha^+_i, \alpha^-_i]
\; \; . \eeq{betbn}

The boundary conditions are precisely those of Ref. \cite{kmw}.  They have
been rederived here to properly account for any singularities in arising
from contact terms in the equations of motion.  These contact terms
when properly regulated do not affect the boundary conditions.

We now construct an approximate solution of the equations of motion in the
forward light cone.  We do this by expanding around the solution which is 
a pure two dimensional gauge transform of vacuum which is closest to
$\alpha^+ + \alpha^-$.  To do this, we introduce the projectile
and target source charge
per unit area at a reference rapidity $y$ as
\beq
        q^\pm(y,x_\perp) = \pm \int_y^{\pm y_{cm}} dy^\prime \rho 
(y^\prime, x_\perp)
\eeq{totch}
and
\beq
        \eta^\pm(y,x_\perp)
 = \pm \int_y^{\pm y_{cm}} dy^\prime \Lambda (y^\prime,
x_\perp)
\eeq{etapm}
Note that
\beq
\langle q^\pm_a(y,x_\perp) q^\pm_b(y,x_\perp^\prime)\rangle
= \chi^\pm(y) \delta_{a,b} \delta^2(x_\perp-x_\perp^\prime)
\eeq{corr}
in terms of $\chi^\pm(y)$ defined as in (\ref{chi}).

By direct computation, as in \cite{kmw}
\beq
        \alpha_i^\pm = \nabla_i \eta^\pm - {i \over 2} [\eta^\pm, \nabla_i
\eta^\pm ]
\eeq{alphofeta}
and
\beq
        \eta^\pm = g^2 {1 \over {\nabla^2_\perp}} q^\pm
\eeq{etofq}

The sum of $\alpha^+ + \alpha^-$ can be written as a pure two dimensional 
gauge transform of vacuum plus a correction as
\beq
        \alpha^+_i + \alpha^-_i = \alpha_i^0 + \delta \alpha_i^0
\eeq{sumalph}
where
\beq
        \alpha_i^0 = \nabla_i (\eta^+ + \eta^-) - {i \over 2}
[\eta^+ + \eta^-, \nabla_i (\eta^+ + \eta^-)]
\eeq{alph0}
and where
\beq
        \delta \alpha^0_i = {i \over 2} \{ [\eta^-, \nabla_i \eta^+]
+ [\eta^+, \nabla_i \eta^-] \}
\eeq{delalph}
This decomposition into a gauge transform of the
vacuum is accurate up to and 
including order $g^4$.

Now we expand $\alpha_i^3 = \alpha^0_i + \delta \alpha^3_i (\tau, x_\perp)$.
Both $\delta \alpha^3 $ and $\beta $ are the small fluctuation
fields corresponding to radiation.  We find that $\delta \alpha^3$ and
$\beta$ solve exactly the same equations as were incorrectly derived
in Ref. \cite{kmw}.  So even though the original derivation was incorrect,
the final result remains  fortunately valid.

In Eqn. 42 \cite{kmw}, a factor of $2\pi $ was however 
omitted, and as well in Eqns.
45, 47, 49 and 50.  In addition, in going from the first of Eqns. 49
to the second, a factor of 1/2 from the trace was omitted.
  
The final result corrected for the above factors and generalized
to include the source of hard gluons is
\begin{eqnarray}
        {{dN} \over {dy d^2k_\perp}} & = & \pi R^2 {{2g^6 \chi^+(y)\chi^-(y)} 
\over {(2\pi)^3}}
{{N_c(N_c^2-1)} \over {k_\perp^2}} \int {{d^2q_\perp} \over {(2\pi )^2}}
{1 \over {q_\perp^2(\vqp-\vkp)^2}}\nonumber \\
   & = & \pi R^2 {{2 g^6 \chi^+(y)\chi^-(y)} 
\over {(2\pi)^4}} {{N_c(N_c^2-1)}
\over k_\perp^4} \call(k_\perp,\lambda)
\; \; .\end{eqnarray}
The $\vqp=0$ and $\vqp=\vkp$ divergences arise in the above classical
derivation because of the neglect of correlations in the sources
ensemble. A finite logarithmic factor , $\call(k_\perp,\lambda)$, is
obtained only if we include a finite color neutralization correlation
scale, $\lambda$.  

This scale arises from dynamical screening effects and may be seen in models
such as the onium valence quark model of Kovchegov\cite{yuri} as developed in
\cite{rischyuri}.  In the classical calculation, this cutoff appears after
averaging over various values of the background charge density.\cite{jamil} The
cutoff scale turns out to be $\lambda \sim \alpha \surd \chi$.  Below this
cutoff scale, the factors of $1/k_\perp^2$ moderate and become of order
$\ln(k_\perp)$.  This cutoff scale acts somewhat like a Debye mass, although
this is not quite the case since the logarithmic dependence implies power law
fall off in coordinate space whereas a Debye mass corresponds to exponential
decay.  In any case, for evaluating $\call(k_\perp,\lambda)$ at $k_\perp >>
\lambda $ the precise form of the cutoff is unimportant, only that the
$1/k_\perp^2$ singularities in the integrand are tempered at some scale.  This
is because logarithmically divergent integrals are insensitive in leading order
to the precise form of the cutoff.  The generic form of the logarithmic factor
is therefore expected to be of the form
\begin{eqnarray}
\call(k_\perp,\lambda,y)&=& k_\perp^2 \int {{d^2\vqp} \over {2\pi}}
{ {\cal F}(y,q_\perp^2){\cal F}(y,(\vqp-\vkp)^2)
 \over {q_\perp^2 (\vqp-\vkp)^2}}
\end{eqnarray}
where ${\cal F}$ is a suitable form factor.
In \cite{gunion} a  dipole  form factor was considered.
A gauge invariant screening mass was considered 
 in \cite{gyuwang}. Such dipole form factors lead to 
\begin{eqnarray}
\call(k_\perp,\lambda)&=& k_\perp^2\int {{d^2\vqp} \over {2\pi}}
{ 
1 \over {(q_\perp^2+\lambda^2)((\vqp-\vkp)^2+\lambda^2)}}
\approx \log(k_\perp^2/\lambda^2)
\; \; , \end{eqnarray}
where the logarithmic form
is remarkably accurate  for $k_\perp/\lambda > 2$.
A finite but nonlogarithmic form of $L$ can also arise if other
functional forms for the form factors are considered as in
\cite{yuri,rischyuri}.

 It is also important to stress  that in any case, the above classical
derivation neglected nonlinearities that can be expected to distort
strongly the above perturbative solution in the $k_\perp^2< \alpha^2
\chi$ region. Thus, the solution should not be extended below
$\lambda\sim\alpha \surd\chi$ in any case.  In future studies, it will
be important to investigate  just how the 
full nonlinear Yang-Mills equations regulate these infrared
divergences.

\subsection{Classical Color Current Fluctuations}

For two colliding nuclei the effective classical source
current for mid-rapidity gluons is assumed to be
\beq
j^{\mu}_a(x) = \delta^{\mu+} \delta(x^-)q^+_a(0,\vxp)
+ \delta^{\mu-} \delta(x^+)q^-_a(0,\vxp)
\eeq{2j}
where $\langle q^\pm\rangle =0$ but the ensemble
averaged  squared color charge density of each of the $N_c^2-1$ components
is given by $\chi^\pm(0)$ as in (\ref{corr}).

In ref.\cite{yuri,rischyuri}, $\chi$ was estimated using
the valence quark density  and with the classical
color density interpreted as a
color transition density associated with the  radiation of
a color $a$ gluon
\beq
q^a(\vxp)= \sum_{n=1}^N (T^a_n)_{c^\prime,c} \delta^2(\vxp -{\vxp}_n)
\eeq{jy}
where the sum is over the valence quarks, and $T^a_n$ is a generator
of dimension $d_n$ appropriate for parton $n$.
In this interpretation,  products of color densities
involve matrix multiplication and the ensemble average leads to a trace
associated with averaging over 
all initial colors of the valence partons and a summing over all final
colors. Therefore
\beq
\langle q_i^a(\vxp)\rangle =0
\eeq{zero}
since  $Tr T^a=0$ in any representation while 
\beq
\langle q^a(\vxp)q^b(\vxp^\prime)\rangle =
\sum_{n=1}^N \frac{1}{d_n}Tr(T^a_n T^b_n) n(\vxp)\delta^2(\vxp -{\vxp}^\prime)
\eeq{r2}
where $n(\vxp)=\langle \delta(\vxp-\vxp(n))\rangle$ is the transverse density
of partons
of type $n$. From now on we assume identical projectile and target
combinations and fix $y=0$ so that we can drop the distinction between $\pm$
sources and the rapidity variable.

Taking into account both the valence quark and hard gluon
contributions in the nuclear cylinder approximation used in \cite{kmw},
the relevant $\chi=\chi^\pm(0)$ parameter is therefore given by
\beq
\chi = 
\frac{1}{\pi R^2}\left(\frac{N_q}{2N_c} + \frac{N_c\; N_g}{N_c^2-1}
\right)
=\frac{1}{\pi R^2}(C_F N_q + C_A  N_g)/d_A
\eeq{chi0}
where the transverse density
of quarks is $n_q(\vxp)=N_q/\pi R^2$ and
the gluon transverse density is $n_g(\vxp)=N_g/\pi R^2$,


Because this interpretation allows for complex color (transition) densities
that do not arise in the  classical limit,  it is useful
to show that it can also be derived from
a more conventional classical Yang-Mills treatment.
For that purpose we use the Wong formulation of classical YM kinetic theory
\cite{wong}.
In that formulation, the parton phase space is enlarged to incorporate
a classical charge vector $\Lambda^a(\tau)$ in addition to the
usual $(x^\mu(\tau),p^\mu(\tau)=m u^\mu(\tau))$ phase space coordinates.
The phase space density, $f$, obeys the Liouville equation 
\beq
\frac{d}{d\tau}f(x(\tau),p(\tau),\Lambda(\tau))=0
\eeq{louis}
with $dx^\mu/d\tau= u^\mu$ and 
\beqar
m\frac{du^\mu}{d\tau} &=& gu_\nu F^{\mu\nu}_a \Lambda^a \nonumber\\
\frac{d\Lambda^a}{d\tau}&=& gf^{abc}u_\mu A^{\mu b} \Lambda^c
= -i \left({\cal T}^b\theta^b\right)_{ac}\Lambda^c
\; \; . \eeqar{wong} 
where $({\cal T}^b)_{ac}=if^{abc}$ are the generators in the adjoint 
representation and $\theta^b=g u_\mu A^{\mu b}$. 
The color current $gj^{\mu a}(x)$ in this
kinetic theory is computed via
\beq
j^{\mu a}(x)= \int d\tau u^\mu(\tau)
\Lambda^a(\tau)\delta^4(x-x(\tau))
\eeq{wongj}
The color charge vector precesses around the local $A^{\mu a}$
field but its magnitude remains constant. Its length is fixed
by the specified color Casimir $C_2=\sum_a \Lambda_a^2$.
In the ultra-relativistic case with $p^z/p_0\approx 1$,
the current reduces to eq.(\ref{2j})
with the transverse density
\beq
q^a(\vxp)=  \sum_{n} U_{ac} \Lambda^c_n(\tau_0)
 \delta^2(\vxp -{\vxp}_n)
\eeq{wongjt}
where the unitary 
$U=P\exp\{-ig\int_0^1 ds {\cal T}^b u_\mu A^{\mu b}(x(s))\}$ 
 accounts for the color
precession along the parton trajectory.
The ensemble average in this formulation involves an integration over 
the initial colors $\Lambda^a_n(\tau_0)$ 
with a measure
\beq
d\Lambda_n \propto \prod_{c=1}^{d_A} d\Lambda^c_n
\delta(\Lambda^a_n\Lambda^a_n-C_{2 n})
\eeq{meas}
normalized such that $\int d\Lambda_n=1$ and thus
\beq
\int d\Lambda_n 
\Lambda^a_n \Lambda^b_n
= \delta^{ab} C_{2n}/d_A
\eeq{lamint}
Because $U$ is unitary,
this leads to the same expression for the color charge squared correlation
parameter $\chi$ as eq.(\ref{chi0}).

\subsection{Yang-Mills Radiation Distribution}

Inserting the above expression for $\chi$ into the classical formula for
radiation, we obtain
\beqar
\frac{dN}{dyd^2k_\perp}&=& 
\frac{1}{\pi R^2}\left(\frac{N_q}{2N_c} + \frac{N_c\; N_g}{N_c^2-1}
\right)^2
\frac{2g^6}{(2\pi)^3}
\frac{N_c(N_c^2-1)}{k_\perp^2}\int\frac{d^2\vqp}{(2\pi)^2}
\frac{1}{q_\perp^2(\vqp-\vkp)^2}\nonumber\\
&=&  \frac{1}{\pi R^2}\left(C_F N_q + C_A  N_g)
\right)^2 \frac{1}{d_A}
\frac{2g^6N_c}{(2\pi)^4}
\frac{1}{k_\perp^4}
\call(k_\perp,\lambda)
\eeqar{new}
If only valence quarks are included then this reduces to
\beqar
\frac{dN}{dyd^2k_\perp}&=& 
  \frac{N_q^2}{\pi R^2}
\left(\frac{2g^6 N_c}{(2\pi)^4}\right)
\left(\frac{C_F^2}{d_A}\right)_{el}\frac{1}{k_\perp^4}
\call(k_\perp,\lambda)
\eeqar{newq}
In the opposite limit, if only  hard glue 
is included, 
  the radiation distribution reduces to
\beqar
\frac{dN}{dyd^2k_\perp}&=& 
 \frac{N_g^2}{\pi R^2}
\left(\frac{2g^6 N_c}{(2\pi)^4}\right)
\left(\frac{C_A^2}{d_A}\right)_{el}\frac{1}{k_\perp^4}
\call(k_\perp,\lambda)
\eeqar{newg}

Note that the color factor in the second brackets marked $el$
is that associated with the
elastic scattering of two partons
\beq
C^{el}_{nm}=\left(\frac{1}{d_n}Tr T^a_nT^b_n \right)\; 
\left(\frac{1}{d_m}Tr T^a_mT^b_m \right)
= \frac{C_{2n} C_{2m}}{d_A}
\;\; , \eeq{cel}
so that $C^{el}=2/9,9/8$ for $qq,gg$.
The elastic Rutherford cross section is in this approximation
\beq
\sigma^{el}_{nm}=\frac{g^4 C^{el}_{nm}}{(2\pi)^2}
\int \frac{d^2 q_\perp}{q_\perp^4} =\int dt \frac{\pi\alpha^2}{t^2} \frac{
4 C_{2n}C_{2m}}{d_A}
\; \; . \eeq{el}
The infrared divergence is regulated by the color screening scale
$\lambda$ or  form
factors as in \cite{gunion}.

The  geometrical Glauber factor in Eqns. ( \ref{newq},\ref{newg})
counts the average
number of binary
parton-parton collisions per unit area in $b=0$ collisions of cylindrical
nuclei. More generally,
\beq
T^{nm}_{AB}({\bf b})= 
\frac{1}{\sigma_{nm}}\int d^3 x \rho_{n/A}(\vx) \int dz_B\; \sigma_{nm}
\rho_{m/B}(\vxp-{\bf b},z_B)
\eeq{tab}
For $b=0$ collisions of cylindrical nuclei this reduces to
\beq
T^{nm}(0)=\frac{N_n N_m}{\pi R^2}
\eeq{tnm}
Therefore we can write
\beqar
\frac{dN}{dyd^2k_\perp}&=& T^{nm}(0) \frac{d\sigma^{nm\rightarrow g}}{dy
d^2 k_\perp}
\eeqar{dsig}
where 
\beqar
\frac{d\sigma^{nm\rightarrow g}}{dy d^2k_\perp}
& = & C^{el}_{nm} 
\left(\frac{2g^6 N_c}{(2\pi)^3}\right)
\frac{1}{k_\perp^2}\int\frac{d^2\vqp}{(2\pi)^2}
\frac{1}{q_\perp^2(\vqp-\vkp)^2}
\; \; . \eeqar{sigrad}

\section{Quantum Radiation}

\subsection{pQCD Bremsstrahlung}
We  compare (\ref{sigrad}) with the quantum radiation
formula derived in \cite{gunion}.
In the $A^+=0$ gauge and for gluon kinematics $k=[xP^+,k_\perp^2/xP^+,\vkp]$
with $x\ll 1$, the three dominant diagrams sum in the small momentum transfer
limit to
\beq
iM({nm\rightarrow g(k,\epsilon,c)})=
[T_n^a,T^c_n]T^a_m \left(\frac{2g^2s}{q_\perp^2}\right)
\left(2g\epsilon_\perp\cdot \left\{ \frac{{\bf k}_\perp}{k_\perp^2}
-\frac{{\bf k}_\perp-{\bf q}_\perp}{({\bf k}_\perp-{\bf q}_\perp)^2}
\right\}\right)
\eeq{mgunion}
Taking the square and averaging over initial and summing over final colors,
one finds that
\beqar
\frac{d\sigma}{d q_\perp^2 dy d^2 k_\perp}&=&
\left(\frac{C^{el}_{nm}4\pi\alpha^2}{t^2}
 \right)\left(
\frac{\alpha N_c}{\pi^2} \frac{q_\perp^2}{
k_\perp^2 ({\bf k}_\perp-{\bf q}_\perp)^2} \right) \nonumber\\
&=&\frac{d\sigma^{el}_{nm}}{dq_\perp^2} \frac{dN}{dyd^2k_\perp}
\; \; . \eeqar{dsiggb}
This is the basic factorized form of the soft QCD
radiation associated with elastic scattering.
Integrating over the elastic momentum transfer $q_\perp$ yields
\beqar
\frac{d\sigma}{dy d^2 k_\perp}&=&\int \frac{d^2\vqp}{\pi}
\left(\frac{C^{el}_{nm}4\pi\alpha^2}{q_\perp^4}
 \right)\left(
\frac{\alpha N_c}{\pi^2} \frac{q_\perp^2}{
k_\perp^2 ({\bf k}_\perp-{\bf q}_\perp)^2} \right) \nonumber\\
&=& 
 C^{el}_{nm}\frac{2g^6 N_c}{(2\pi)^3} 
\frac{1}{k_\perp^2}\int\frac{d^2\vqp}{(2\pi)^2}
 \frac{1}{
q_\perp^2 ({\bf k}_\perp-{\bf q}_\perp)^2}
\eeqar{dsiggb2}
This is exactly the same as the classical result in (\ref{sigrad}).

In  Ref. \cite{gunion} the $\pi\pi\rightarrow g$ cross section
was computed taking a dipole form factor into account
with the result
\beqar
\frac{d\sigma^{\pi\pi\rightarrow g}}{dy dk_\perp^2}
= 
\left(\frac{C_A\alpha^3}{\pi^2 k_\perp^2 }\right)
\int\frac{d^2\vqp}{(2\pi)^2}
\frac{2^2F_\pi(q_\perp^2)
F_\pi((\vqp-\vkp)^2)}{q_\perp^2(\vqp-\vkp)^2}
\eeqar{sigradpi}
where
\beq
F_\pi(q^2)=\frac{4 q^2}{4 q^2 + m_\rho^2}
\eeq{fpi}
Again we  can read off  the
elementary $qq\rightarrow g$ cross section 
by dividing by the number of parton pairs $N_q^2=4$
in this reaction and 
 neglecting interference by setting $F_\pi=1$.
This leads  to 
\beqar
\frac{d\sigma^{qq\rightarrow g}}{dy dk_\perp^2}
= \frac{1}{4}
\left(\frac{2 g^6 N_c}{(2\pi)^3}\right)
\frac{1}{k_\perp^2}\int\frac{d^2\vqp}{(2\pi)^2}
\frac{1}{q_\perp^2(\vqp-\vkp)^2}
\eeqar{sigrad2}
where the first factor 1/4 is just the large $N_c$ limit of
$C^{el}_{qq}\rightarrow 1/4$ used implicitly in eq.(17) of \cite{gunion}.

\subsection{Comparison with GLR  formula}

It is also of interest to connect the classical YM formula with the
$pp\rightarrow g$ formula of Gribov, Levin, and Ryskin (GLR)\cite{gribov}and
used recently in ref.\cite{esk}to compute mid-rapidity gluon production at LHC
energies: \beqar \frac{d\sigma}{dyd^2{\bf k}_\perp} &=& K_N \frac{\alpha
  N_c}{\pi^2 k_\perp^2} \int d^2{\bf q_\perp} \frac{f(x_1,q_\perp^2)
  f(x_2,({\bf k}_\perp -{\bf q_\perp})^2)}{q_\perp^2 ( {\bf k}_\perp -{\bf
    q_\perp})^2} \; \; ,\eeqar{glr} where \beq f(x,Q^2)= \frac{d}{d\log Q^2}
xG(x,Q^2) \eeq{fq} and \beq x_1\approx x_2\approx x_\perp\equiv
k_\perp/\sqrt{s} \eeq{x} are fractional momenta which are assumed to be small.
In this relation, the radiation resulting from 
the fusion of two off-shell $y_1\sim
y_2\sim 0$ gluons is estimated. Unfortunately, there is variation in the
literature as to the magnitude of the factor $K_N$ \cite{gribov}. This is
partly due to variations in the definition of $f(x,Q^2)$.  We find below that
in order to reproduce the perturbative QCD and classical Yang-Mills result,
eq.(\ref{dsiggb2},\ref{sigrad}), we must take \beq K_N=
\frac{(2\pi)^2}{N^2_c-1}\approx 5 \; \; . \eeq{kappa} From private
communication with E. Levin, this factor is required if $f$ is defined as in
ref.\cite{esk} via (\ref{fq}).  This implies that the results quoted for the
BFKL contribution to minijets in \cite{esk} taking $K_N=1$ are approximately a
factor 5 too small. With the value in eq.(\ref{kappa}), the BFKL and
conventional mini-jet rates would coincide more closely. In section 3.3 we
argue that at least in the asymptotic domain these distributions should in fact
coincide over a range of $k_\perp\sim\surd\chi$.

To compare to the GLR formula with the classical result, 
we  approximate the $Q^2$ evolution using 
DGLAP evolution\cite{dglap}
\beq
f(x,Q^2)= \frac{d xG(x,Q^2)}{d \log Q^2}\approx \frac{\alpha}{2\pi}
\int_x^1 \frac{dx^\prime}{x^\prime}\; G(x^\prime,Q^2) \; 
x P_{g\rightarrow g}
(x/x^\prime)
\; \; .
\eeq{altpar}
In the small $x$ semi-classical domain
\beq
\; \; P_{gg}(x)\approx 2N_c/x
\eeq{split}
Therefore,  we have the approximate relation at high $Q^2$
\beq
f(x,Q^2)\approx \frac{ \alpha N_c}{\pi} \int_x^1 \;dx^\prime \;G(x^\prime,Q^2)
= \frac{ \alpha N_c}{\pi}
 N_g(x,Q^2)
\eeq{fapp}
Consequently, from (\ref{fq},\ref{fapp})
\begin{eqnarray}
\frac{d\sigma}{dyd^2{\bf k}_\perp} &=& 
K_N \frac{\alpha N_c}{\pi^2 k_\perp^2}
\int d^2{\bf q_\perp} \frac{d\; xG}{dq_\perp^2} \;
\left(\frac{d\;xG}{dq_\perp^{\prime 2}}\right)_{(
{\bf k}_\perp -{\bf q_\perp})^2}
 \nonumber\\[2ex]
&\approx&  K_N \frac{\alpha N_c}{\pi^2 k_\perp^2}
\frac{\alpha^2 N_c^2}{\pi^2}
\int d^2{\bf q_\perp} \frac{N_g(x_1,q_\perp^2) 
N_g(x_2,({\bf k}_\perp -{\bf q_\perp})^2)}{q_\perp^2 (
{\bf k}_\perp -{\bf q_\perp})^2}\nonumber\\[2ex]
\eeqar{glr2}
Equation (\ref{glr2}) reduces to the classical expression 
(\ref{newg}) if we approximate
the integral by factoring out  the integrated gluon numbers
at the average scale, $\sim k_\perp^2$, divide by $\pi R^2$, and take
 the normalization
factor $K_N$ from (\ref{kappa}).

\subsection{Matching \protect{$2\rightarrow 3$} to \protect{$2\rightarrow 2$}}

Up to this point, we have shown that the classical and quantum bremsstrahlung
formulas agree for the $2\rightarrow 3$ process up to specific form factors.
The problem addressed in this section is the relationship between the
bremsstrahlung spectrum and the mini-jet spectrum based on the pQCD factorized
$2\rightarrow 2$ processes.  Recall\cite{mini} the factorized differential
cross section for two gluon jet production with transverse momenta , $\pm\vkp$,
and rapidities $y_1$ and $y_2$, is given by
 \beq {{d\sigma^{AB\rightarrow
      g_1g_2 X}} \over {dy_1 d y_2 d^2\vkp}} = K\;x_1 G_{A}(x_1, k_\perp^2) x_2
G_{B}(x_2, k_\perp^2) \frac{1}{\pi} \frac{d\sigma^{gg\rightarrow gg}}{dt} \; \;
, \eeq{partscat} 
where $x_1=x_\perp(\exp(y_1)+\exp(y_2))$ and
$x_2=x_\perp(\exp(-y_1)+\exp(-y_2))$, with $x_\perp=k_\perp/\surd s$, and where
the pQCD $gg\rightarrow gg $ cross section for scattering with
$t=-k_\perp^2(1+\exp(y_2-y_1))$ and $y_2-y_1=y$ is given by \beqar
\frac{d\sigma^{gg}}{dt}&=& 
C^{el}_{gg}
\frac{4\pi\alpha^2}{k_\perp^4}\frac{(1+e^y+e^{-y})^3}{(e^{y/2}+e^{-y/2})^6}
\; \; . \eeqar{ggqcd}
This 
reduces to the naive Rutherford expression eq.(\ref{el})
only if the unobserved gluon has a rapidity $|y|\gsim 3$.
For $|y|\lsim 1$, the exact form (\ref{ggqcd})
is $27/64\approx 0.42$ smaller than the Rutherford approximation.

We concentrate  here only on the dominant gluon-gluon contribution
for symmetric systems, $A+A$, with $G=G_A$.
The inclusive gluon jet production cross section is obtained by integrating
over $y_2$ with $y_1=y$ and $k_\perp$ fixed. For an observed
midrapidity gluon with  $y=0$, $-y^*<y_2<y^*$,
where  $\exp(-y^*)=x_\perp/(1-x_\perp)$, we must evaluate
\beqar
I(x_\perp,k_\perp^2)&=&\int_{-y^*}^{y^*} dy_2  \;x_1 G(x_1, k_\perp^2)
x_2 G(x_2, k_\perp^2) 
\frac{(1+e^{y_2}+e^{-{y_2}})^3}{(e^{y_2/2}+e^{-y_2/2})^6}
\; \; . \eeqar{intex}
In the Rutherford approximation, implicit in the
classical approximation, we neglect the $y$ dependence of
(\ref{ggqcd}) and therefore approximate $I$ by
\beqar
I_{R}(x_\perp,k_\perp^2)&=&\int_{-y^*}^{y^*} dy_2  \;x_1 G(x_1, k_\perp^2)
x_2 G(x_2, k_\perp^2) \\ \nonumber
&\approx& 2 x_\perp G(x_\perp ,k_\perp^2)
\int_{x_\perp}^1 dx_2
 G(x_2,k_\perp^2)
\\ \nonumber
&\approx&
2 x_\perp G(x_\perp ,k_\perp^2) N_g(x_\perp,k^2_\perp)
\; \; . \eeqar{app1}
For $xG \propto x^{-\delta}(1-x)^\gamma$ with $\delta\sim 0.2,\gamma\sim 8.5$,
as  HERA data\cite{zeus,h1} indicate in the moderate 
$Q^2\sim 5$ GeV$^2$ range,
the  last approximation to $I_R$ is found to agree remarkably
within  10\% of the numerical
integral of the first line as long as  $x_\perp\lsim 0.01$. 
However,  for $k_\perp^2=5$ GeV$^2$,
the neglect of the rapidity dependence 
of $d\sigma_{gg}/dt$ in the Rutherford approximation 
leads $I_R$ to overestimate
 $I$ by $\sim 55\%$ at RHIC energies ($x_\perp\sim 0.01$) and 
by $\sim 34\%$ even at LHC energies  $x_\perp\sim 0.001$.
This is  due to the factor $\sim 2$ suppression of the pQCD rate in the 
 $|y_1-y_2|<1$ range. On the other hand, next to leading order 
corrections modify (\ref{partscat}) by a factor
 $K\sim 2$ in any case, and the next to leading order corrections to the
 classical formula are not yet known.
Since neither the mini-jet nor the classical radiation
can be determined at present to better than $\sim 50\%$
accuracy,  the following
 simplified Rutherford formula for the single inclusive
pQCD mini-jet distribution is adequate:
\beq
        {{d\sigma} \over {dy dt}} \approx 2N_g(x,t) xG(x,t) 
\left({{d\sigma_{gg}^{el}} \over {dt}}\right)_{R}
\eeq{intcrs}
or
\beq
        {{dN} \over {dydt}} \approx  {{2N_g(x_\perp,t)} \over {\pi R^2}}
        x_\perp G(x_\perp,t) 
\left({{d\sigma_{gg}^{el}} \over {dt}}\right)_R
\eeq{dnint}

In order to compare the above mini-jet distribution
with the classical bremsstrahlung result (\ref{newg}),
we  need to replace the   $N_g^2$ factor in (\ref{newg})
by $N_g(x_\perp,q_\perp^2)
N_g(x_\perp,(k_\perp
-q_\perp)^2)$
and move that factor inside the logarithmic integrand.
This generalization is essential since the effective classical source
due to hard gluons depends on the $x_\perp$ and scale of resolution
of the radiated gluon. 
This requires that $k^2_\perp$ be sufficiently large
so that the variation of the structure function 
with  that  scale be small.
In this case, the classical bremsstrahlung formula generalizes  into 
the GLR form (\ref{glr2})
\begin{eqnarray}
        {{dN} \over {dydt}}  
& = & \frac{1}{\pi R^2}~ {4 \alpha N_c}
{{d\sigma^{el}_{gg}} \over {dt}} k_\perp^2 
\int {{d^2\vqp} \over {(2\pi)^2}} 
{{N_g(x_\perp, q_\perp^2)} \over {q_\perp^2}} {{N_g (x_\perp,
(\vkp-\vqp)^2)} \over {(\vkp-\vqp)^2}}
\nonumber \\
       & \approx & {1 \over {\pi R^2}} {8 \alpha N_c}
 N_g(x_\perp,k_\perp^2)
{{d\sigma^{el}_{gg}} \over {dt}}
 \int {{d^2\vqp} \over {(2\pi)^2}} 
{{N_g(x_\perp,q_\perp^2)} \over {q_\perp^2}} \theta(k^2_\perp \gsim q_\perp^2) 
\nonumber \\
       & \approx & {1 \over {\pi R^2}}  ~2 N(x_\perp,k_\perp^2)
{{d\sigma^{el}_{gg}} \over {dt}}
 \int_0^{k_\perp^2} dq_\perp^2 \frac{d}{dq^2_\perp} x_\perp 
G(x_\perp,q_\perp^2) 
\; \; , \eeqar{genclass}
where in the last step we used
the DGLAP evolution (\ref{fapp}). Thus, we recover
the same minijet formula as (\ref{dnint}).

The use of the DGLAP evolution is essential to prove the duality
between  classical bremsstrahlung and the conventional
minijet distributions.
We note  that in order for corrections to the classical result to remain 
small, it is  necessary that 
$\alpha(\chi) \ln(k_\perp^2/\alpha^2 \chi) \ll 1$.
Recall that in the classical analysis, $\alpha$ is always to be evaluated
at some scale of order $\surd \chi\gg\Lambda_{QCD}$. 
 This requirement is therefore that
$k_\perp \sim \surd \chi$.
  If this is satisfied, then the formulae should agree in the $x_\perp\ll 1$
regime.

We see therefore that all the formulae used for hard scattering agree with
the classical result in the range of momentum $\alpha^2 \chi \ll k_\perp^2 \lsim
\chi$.  This range of momenta is outside the typical
scale $k_\perp^2 \sim \alpha^2 \chi$ on which 
non-trivial behaviour
of  the transverse momentum distributions is expected on account of
screening. In the region of smaller $k_\perp$, the full nonlinearity
of the Yang Mills equations must be taken into account.
At large
 $k_\perp>\surd \chi$,  the hard scattering pQCD formula  properly
sums up higher order DGLAP corrections to the classical formula.
 It is important  that there is a range of momenta where the
classical and hard scattering results match 
at the level of $\sim 50\%$.

\section{Estimate of \protect{$\chi(A,s,Q^2)$}}

We turn finally to the estimate of the McLerran-Venugopalan scale
density $\chi$ in the  range of   $A$ and $s$ 
in future RHIC and LHC experiments.

\subsection{Valence Quark Contribution} 

The initial assumption in \cite{mcven} and
further developed  in \cite{yuri,rischyuri} was that for $A\gg 1$,
the valence  quarks could provide a very high density of 
 hard color source partons for which recoil effects are negligible
and thus treated classically.
In the nuclear cylinder approximation, the transverse  density
of valence quarks is simply
\beq
n_q=\frac{N_c A}{\pi R^2} = \frac{N_c A^{1/3}}{\pi r_0^2}
\eeq{valq}
where $r_0=1.18$ fm. Since  each quark contributes with a color
factor $  C_F/d_A=
{1}/{2N_c}$ the valence quark contribution to the color charge squared
density 
\beq
\chi_{val}= \frac{A^{1/3}}{2\pi r_0^2}
=(A^{1/6}\; 0.07\;{\rm GeV})^2 \lsim  \Lambda^2_{QCD}
\; \; , \eeq{val}
where the bound arises because only $A<200$ beams will be available.
One would need astronomical $A\sim 10^6$ to reach $\surd \chi_{val}=1$ GeV
because of the extremely slow $A^{1/6}$ growth. Thus, 
the valence quark contribution is in practice too
dilute to contribute into the perturbative domain.
Fortunately, the non-Abelian Weizs\"acker-Williams
field has a very large number of ``semi-hard'' gluons with $x\gsim x_\perp$
in the $s\rightarrow \infty$ limit. The question then is
how large does $s$ have to get in order
to push  $\chi$ into the perturbative regime.

\subsection{ The Hard Gluon Source}

 The number of hard gluons that can act as a classical source
for midrapidity gluons
depends not only on $x_\perp$ but also the resolution scale $Q$
\beq
N_g(x_\perp,Q^2)=\int_{x_\perp}^1 G(x,Q^2)dx
\; \; . \eeq{gl}

Each gluon contributes to $\chi$ with a color factor $C_A/d_A=3/8$.
In order to get an upper bound, we will neglect possible nuclear glue shadowing
and assume that $N_g\propto A$. 
There could be some suppression of the low $x$ gluon number in nuclei due
to shadowing as observed for nuclear quarks. 
In the McLerran-Venugopalan model, only a mild logarithmic
$\sim \ln(A)$ modification of $N_g \propto A$ is expected.
Including valence and sea quarks and antiquarks 
as well as gluons leads then to 
\beqar
\chi(A,s,Q^2)&\approx& \frac{A^{1/3}}{\pi\; r_0^2}
 \int^1_{x_0}dx \;\left(\frac{1}{6}q(x,Q^2)+\frac{3}{8}  g(x,Q^2)\right) 
\; \; . \eeqar{mus}
The lower bound, $x_0$, 
is determined up to a factor of $\sim 2$
by the minimum momentum fraction
needed to justify the neglect of recoil associated with the radiation of
a midrapidity gluon with a given $k_\perp$. 
In the estimate below we vary that bound between $x_0=x_\perp$
 and $2x_\perp$. For $xg(x)\propto 1/x^\delta$ with $\delta\sim 0.2-0.3$,
this leads to an uncertainty,
$\delta\chi/\chi\sim \delta$, well within the present overall normalization
uncertainties of the small $x$ gluon structure function.

In order to compute $\chi$ we use the
GRV95 NLO($\overline{MS}$) parametrization\cite{grv95} of the nucleon 
structure functions.
Figure 1 shows how the parametrization of the gluon structure
compares to preliminary ``data'' at $Q^2=7$ GeV$^2$
from HERA\cite{zeus,h1} obtained via a DGLAP analysis of the scaling violations
from $F_2(x,Q^2)$.
Also shown is the BFKL-like parametrization of the gluon structure
used in \cite{esk} for comparison.
Both the  GRV95 and the BFKL parametrizations significantly overestimate
the moderate $Q^2$ data of interest here at $x<10^{-3}$. 
The preliminary $Q^2=20$ data from H1\cite{h1}
(not shown Fig.1) also lie below the GRV95 parametrization.
For our purposes, it is only important that  the use of  GRV95
and the neglect of
gluon shadowing should lead to a reasonable
upper bound on $\chi$.

As discussed in the previous sections, the classical regime extends
up to $k_\perp \lsim Q_{YM}$ where
\beq 
Q_{YM}(A,s)=\surd \chi(A,s,Q^2_{YM})
\sim \;(1 \; {\rm GeV})\; \left(\frac{A}{200}\right)^{1/6} 
\; \left(\frac{10^{-4}}{x_\perp}\right)^{\delta/2} 
\; \; , \eeq{qym}
and $\delta\sim 0.2-0.3$. In principle,
this  must be determined self-consistently given the scale dependence
of the glue. In practice, as shown below, $Q_{YM}$ is only weakly dependent
of the reference scale if its above $\sim 2$ GeV$^2$.
The approximate formula for $Q_{YM}$ summarizes
the numerical results  below.

In Figure 2, $Q_{YM}$ is plotted for $\sqrt{s}=0.2,6.5,100$ ATeV for heavy
nuclear beams with $A=200$ as a function of the scale $Q$ with which the GRV95
structure functions are evaluated in (\ref{mus}).  The upper solid curves for
each energy correspond to (\ref{mus}) with $x_0=x_\perp$.  The lower curves are obtained by
increasing the lower cutoff from $x_\perp$ to $2 x_\perp$. The two long dashed
curves at the bottom show the contribution of only the valence quarks to
$Q_{YM}$ at $\surd s=0.2,100$ ATeV using $x_0=x_\perp$.  The curves show that
the hard nuclear glue dominates for finite nuclei at all collider energies.
Note also that $Q_{YM}$ is remarkably independent of the reference $Q$ scale
because of a compensation of two competing effects.  The increase of $xG$ with
$Q$ is compensated by its decrease with increasing value of the minimum hard
fraction $x_0\sim Q/\surd s$ contributing to the classical source.

At RHIC energies, the boundary of the classical regime remains rather low
($\lsim 500$ MeV) because the relevant $x$ range, $x_\perp>0.005$, is not very
small. By LHC energies, on the other hand, gluons down to $x_\perp\sim 0.0001$
can contribute, and the classical Yang-Mills scale increases to $Q_{YM}\sim 1$
GeV.  Note that to double the $Q_{YM}$ scale at a fixed energy would require an
increase of $A$ to $2^6\times 200$ if shadowing can be ignored.  To double the
value of $Q_{YM}$ at fixed $A$ requires decreasing $x_\perp$ by a factor
$2^{-10}\sim 10^{-3}$.  Although asymptotically the scale of $Q_{YM}$ becomes
arbitrarily large, this asymptotic behaviour is approached slowly.  We conclude
that at RHIC energies the classical Yang-Mills radiation dynamics is likely to
modify mainly the nonlinear, nonperturbative beam jet regime.  In that regime
the perturbative analysis must certainly be extended into the full nonlinear
regime via detailed numerical simulations.  By LHC energies it appears that the
classical Yang-Mills radiation begins to overlap into the perturbative minijet
domain with $k_\perp\sim 1$ GeV.

The very small $x$ and very large $A$ limits, where perturbative classical
radiation can be computed, provide an novel calculable theoretical limit.  It
provides qualitatively useful insight at RHIC energies and may be
semi-quantitative already at LHC energies.  In future studies it will be
especially important to extend work with this model into the nonlinear regime
to clarify the mechanisms for color screening in $A+A$ reactions at the lower
$k_\perp\sim \alpha Q_{YM}$ scale. Present estimates for initial conditions in
$A+A$ based on mini-jet pQCD analysis\cite{mini,wang,geiger} vary considerably
because of the necessity to introduce a cutoff scale, $p_0\sim 1-2$ GeV, to
regulate the naive infrared divergent Rutherford rates.  That cutoff has thus
far been estimated either (1) phenomenologically by imposing observed
constraints from extensive $pp,p\bar{p}\rightarrow \pi,K,p,X$ systematics as in
\cite{wang,geiger} or (2) using kinetic theory estimates \cite{muller} which
are sensitive to formation physics effects. One of the great theoretical
advantages of the classical Yang-Mills approach is that the long wavelength
nonlinear dynamics involving pre-asymptotic field configurations can be taken
into account (at least numerically) without invoking kinetic theory or
formation physics assumptions.  In the theoretical $\alpha Q_{YM}\gg 1$ GeV
domain, that physics may be accessible using perturbative techniques. In the
experimentally accessible $\alpha Q_{YM}< 1$ GeV regime, numerical solutions of
the Yang-Mills equations, as for example in \cite{muller2}, are likely to
provide additional insight into that problem.  The classical Yang-Mills
model\cite{mcven,kmw} is one of the practical tools at present to approach the
study of asymptotically high energy reactions, where many unsolved and
interesting theoretical problems remain.\\[2ex]

Acknowledgements: We are grateful to the Institute of Nuclear Theory and Wick
Haxton for supporting the INT-96-3 program where this work was performed.
Numerous useful discussions with K. Eskola, X. Guo, Y. Kovchegov, A. Kovner, J.
Jalilian-Marian, K. Lee, A. Leonidov, E. Levin, A. Makhlin, A. Mueller, D.
Rischke, S. Ritz, R. Venugopalan, and B. Zhang and other participants during
that program are also gratefully acknowledged.  This work was also supported by
the Director, Office of Research, Division of the Office of High Energy and
Nuclear Physics of the Department of energy under contracts DOE-FG02-93ER40764
and DOE-FG02-87ER40328.

\section{Figure Captions}

\begin{flushleft}
Figure 1: The GRV95 NLO\cite{grv95} and BFKL-like\cite{esk} parametrizations of
the gluon structure function, $xG(x,Q^2)$, for $Q^2=7,20 $ GeV$^2$
are compared to {\em preliminary} ZEUS ``data''\cite{zeus} from HERA.\\[2ex]

Figure 2: The classical Yang-Mills scale, $Q_{YM}$, from (\ref{mus})
is shown for $A=200$ nuclei at collider energies $\surd s=0.2,6.5,100$ ATeV
as a function of the reference scale, $Q$, used to evaluate the GRV95\cite{grv95} structure
functions. Upper curves and lower curves for each energy correspond 
to taking the lower cutoff scale $x_0=x_\perp,2x_\perp$, respectively.
The bottom two dashed curves give the valence quark contributions 
at $\surd s=0.2,100$ ATeV.
\end{flushleft}

\newpage
\begin{figure*}[hbt]
\vspace{1.5in}
\hspace{0.5in}\psfig{figure=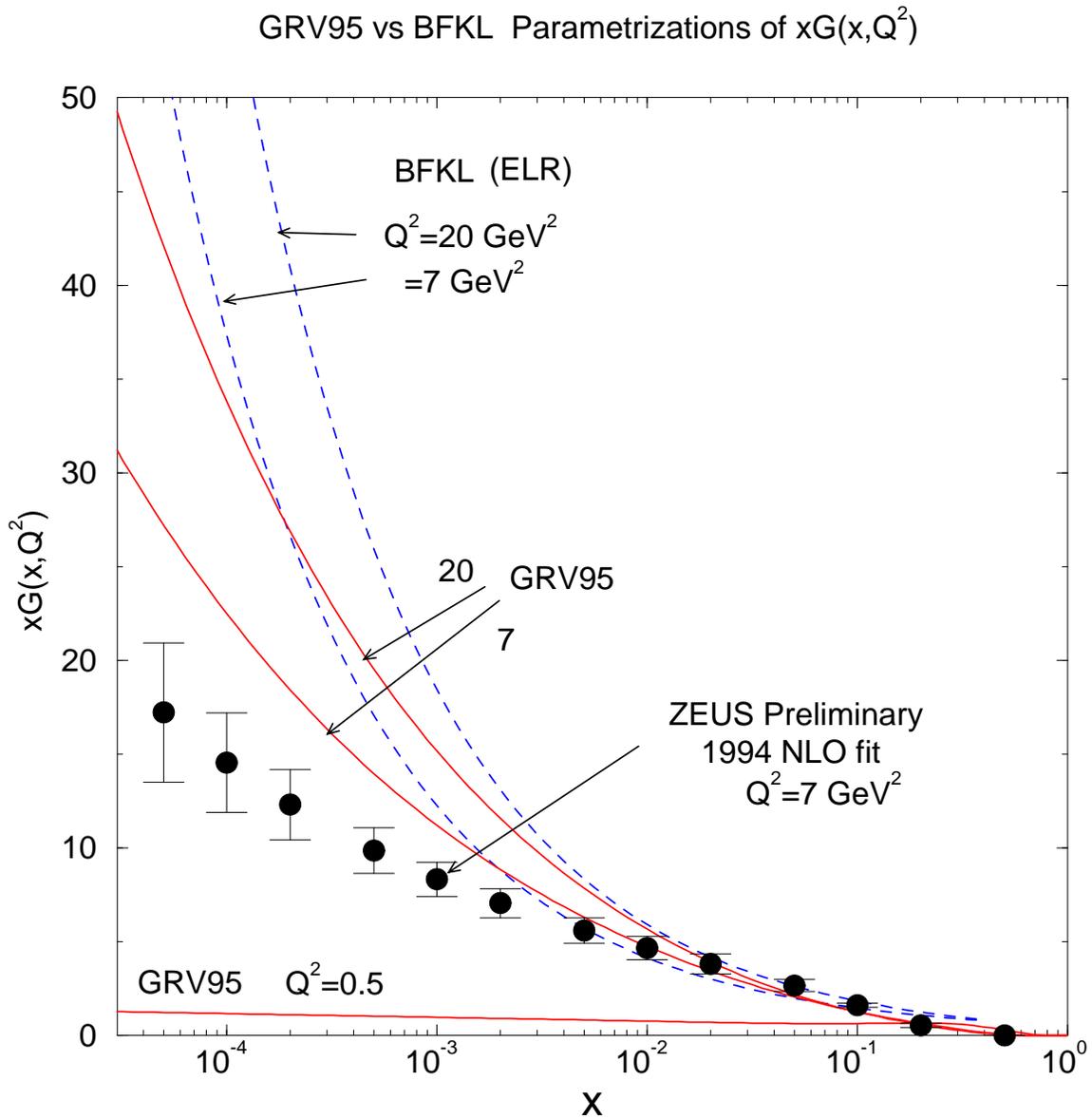,height=5.in,width=5.in,angle=-90} 
\caption{ The GRV95 NLO\protect{\cite{grv95}}
 and BFKL-like\protect{\cite{esk}}
 parametrizations of
the gluon structure function, $xG(x,Q^2)$, for $Q^2=7,20 $ GeV$^2$
are compared to {\em preliminary} ZEUS ``data''\protect{\cite{zeus}}
 from HERA.}
\label{fig1}
\end{figure*}
\newpage
\begin{figure*}[hbt]
\vspace{1.in}
\hspace{0.5in}\psfig{figure=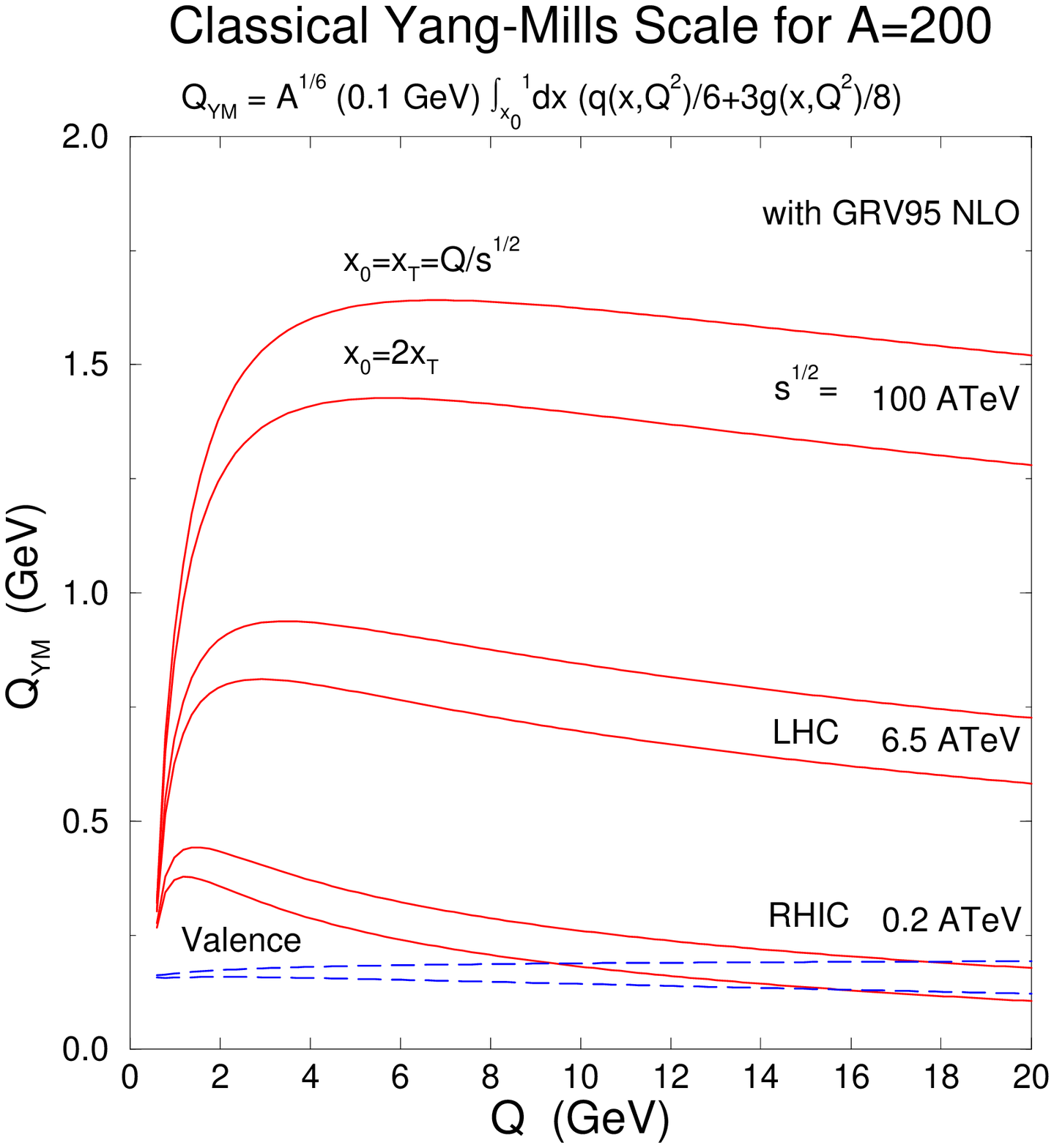,height=5.in,width=5.in,angle=0}
\vspace{1in} 
\caption{ The classical Yang-Mills scale, $Q_{YM}$, from (\protect{\ref{mus}})
is shown for $A=200$ nuclei at collider energies $\surd s=0.2,6.5,100$ ATeV
as a function of the reference scale, $Q$, used to evaluate the 
GRV95\protect{\cite{grv95}} structure
functions. Upper curves and lower curves for each energy correspond 
to taking the lower cutoff scale $x_0=x_\perp,2x_\perp$, respectively.
The bottom two dashed curves give the valence quark contributions 
at $\surd s=0.2,100$ ATeV..}
\label{fig2}
\end{figure*}


\begin{thebibliography}{11}

\bibitem{mcven} Larry McLerran, Raju Venugopalan,
Phys.Rev.D49 (94) 2233 ; 3352 ;
Phys.Rev. D50 (1994) 2225; Phys.Rev.D53:458-475,1996

\bibitem{kmw}Alex Kovner, L. McLerran, H. Weigert,
Phys.Rev.D52 (95) 3809


\bibitem{gunion} J.Gunion and G. Bertsch, PRD 25 (82) 746

\bibitem{gribov} L.V. Gribov, E.M. Levin, M.G. Ryskin ,
Phys.Lett. 100B (1981) 173; 
Phys.Lett. 121B (1983) 65;
Phys.Rept. 100 (1983) 1; 
E.M. Levin and M.G. Ryskin, Phys.Rept. 189 (1990) 267. 


\bibitem{esk} K.J. Eskola, A.V. Leonidov, P.V. Ruuskanen, 
Nucl. Phys. B481 (1996) 704.

\bibitem{mini} K.\ Kajantie, P.V.\ Landshoff, J.\ Lindfors,
Phys.\ Rev.\ Lett.\ 59 (1987) 2527, 
J.P.\ Blaizot, A.H.\ Mueller, 
Nucl.\ Phys.\ B 289 (1987) 847, 
K.J.\ Eskola, K.\ Kajantie, J.\ Lindfors,
Nucl.\ Phys.\ B 323 (1989) 37.

\bibitem{wang} 
X.N.\ Wang and M.\ Gyulassy,  Phys.\ Rev.\ D 44 (1991) 3501;
 D 45 (1992) 844, Phys.\ Rev.\ Lett.\ 68 (1992) 1480.

\bibitem{geiger} K. Geiger. BNL-63762, Jan 1997, 
e-Print Archive: hep-ph/9701226; Phys.Rev.D54:949-988,1996. 
 
\bibitem{qm96} P. Braun-Munzinger, H.J. Specht, R. Stock, and H. St\"ocker, 
eds., Quark Matter '96, Nucl. Phys. A610 (1996) 1c.
\bibitem{jamil}J. Jalilian-Marian, A. Kovner, L. McLerran, H.
Weigert,  hep-ph/9606337.


\bibitem{yuri} Yu.V. Kovchegov, PRD54 (1996) 5463.

\bibitem{rischyuri} Y.V. Kovchegov and D. H. Rischke, CU-TP-824 (1997), 
e-Print Archive: hep-ph/9704201.

\bibitem{bfkl} E.A. Kuraev, L.N. Lipatov, V.S. Fadin, 
 Sov.Phys.JETP 45:199-204,1977;  Ya.Ya. Balitskii, L.N. Lipatov 
Sov.J.Nucl.Phys.28:822-829,1978;

\bibitem{gyuwang} M. Gyulassy and X.N. Wang, 
Nucl. Phys. B420 (1994) 583.

\bibitem{wong} S.K.Wong, Nuovo Cim 65A (1970) 689;
U. Heinz, Ann Phys. (N.Y.) 161 (1985) 48; A.V. Selikov and M. Gyulassy,
Phys. Lett. B316 (1993) 373.

\bibitem{dglap} V.N. Gribov, L.N. Lipatov, 
 Sov.J.Nucl.Phys.15:675-684,1972; G. Altarelli and  G. Parisi,
 Nucl.Phys.B126:298,1977; Yu. Dokshitzer, Sov.Phys.JETP 46 (1977) 1649.


\bibitem{zeus}  S. Ritz private communication;
ZEUS preliminary data presented at
the DURHAM workshop 1996.
\bibitem{h1} S. Aid et al., H1 Collab., Nucl. Phys. B470 (1996) 3 

\bibitem{grv95} M. Gluck, E. Reya, A. Vogt,
Z.Phys.C67 (95) 433


\bibitem{muller} 
K.J. Eskola, B. M\"uller, Xin-Nian Wang, 
Phys.Lett.B374 (1996) 20.

\bibitem{muller2}  T.S. Biro, C. Gong, B. M\"uller, 
A. Trayanov,
Int. J. Mod. Phys. C5 (1994) 113-149; Phys.Rev.D49 (1994) 607, 
5629. 
\end{thebibliography}
\end{document}